\documentclass[
preprint,
 amsmath,amssymb,
 aps,
prb,
floatfix,
]{revtex4-2}

\usepackage{graphicx}
\usepackage{dcolumn}
\usepackage{booktabs, tabularx, ragged2e}
\usepackage{bm}
\usepackage[mathlines]{lineno}
\usepackage{commath}
\usepackage{physics}
\usepackage{color}
\usepackage{mathtools}

\begin{document}

\preprint{APS/123-QED}

\title{Anisotropic spin-split states with canted persistent spin textures in two-dimensional Janus $1T^{'}$ $MXX'$ ($M$ = Mo, W; $X\neq X'$= S, Se, Te) controlled by surface alloying}

\author{Moh. Adhib Ulil Absor}
\affiliation{Departement of Physics, Faculty of Mathematics and Natural Sciences, Universitas Gadjah Mada, Sekip Utara BLS 21 Yogyakarta 55186 Indonesia.}
\email{adib@ugm.ac.id}

\author{Muhammad Arifin}
\affiliation{Departement of Physics, Faculty of Mathematics and Natural Sciences, Universitas Gadjah Mada, Sekip Utara BLS 21 Yogyakarta 55186 Indonesia.}

\author{Iman Santoso}
\affiliation{Departement of Physics, Faculty of Mathematics and Natural Sciences, Universitas Gadjah Mada, Sekip Utara BLS 21 Yogyakarta 55186 Indonesia.}

\author{Harsojo}
\affiliation{Departement of Physics, Faculty of Mathematics and Natural Sciences, Universitas Gadjah Mada, Sekip Utara BLS 21 Yogyakarta 55186 Indonesia.}


\date{\today}

\begin{abstract}
Two-dimensional tungsten-based transition metal dichalcogenides (TMDCs), $MX_{2}$ ($M$: W, Mo; $X$: S, Se, Te) monolayers (MLs) with a $1T'$ structure, serve as significant-gap quantum spin Hall insulators. However, due to the centrosymmetric nature of these crystals, spin degeneracy persists throughout their electronic band structures, limiting their potential for spintronic applications. By modifying the chalcogen ($X$) atoms in the TMDCs ML surface to create a higly stable Janus $MXX'$ MLs structure, we demonstrate through density-functional theory calculations that substantial spin splitting of the electronic states can be achieved. Taking the Janus $1T'$ WSTe ML as a representative case, we identify pronounced anisotropic spin-splitting bands, with maximum spin splittings of 0.14 eV and 0.10 eV occurring at the highest occupied states and lowest unoccupied states, respectively. These significant band splittings give rise to canted persistent spin textures (PST) in the spin polarization, which differ significantly from those in commonly studied PST materials. We demonstrate that this intricate spin splitting and unique spin textures stem from strong in-plane $p-d$ orbital interactions between tungsten (W) and the chalcogen atoms (Te and Se), driven by the reduced symmetry of the crystal's point group. Further analysis using a $\vec{k}\cdot\vec{p}$ model derived from symmetry considerations corroborates the origins of the observed anisotropic spin splitting and canted PST. the spin-split states are highly sensitive to surface imperfections caused by surface alloying effects, such as variations in the chalcogen composition on the monolayer surface. These findings underscore the potential of Janus $1T'$ $MXX'$ MLs as promising candidates for next-generation spintronic devices.
\end{abstract}

\pacs{Valid PACS appear here}
\maketitle

\section{INTRODUCTION}

The interplay between electron spin and orbital degrees of freedom through spin-orbit coupling (SOC) is significantly enhancing the ability to generate, control, and transport spin polarization, opening up new possibilities for spin transport and spin-driven magnetic torque applications beyond conventional spintronic materials \cite{Han2018, Bonell2020}. Topological materials are a major focus in this field, as their key properties often stem from strong SOC combined with band inversions, where topologically protected surface states may enable the generation of coherent, dissipationless spin currents over long distances \cite{Fan2016, Kononov2020}. Three-dimensional (3D) Weyl semimetals (WSMs) stand out due to their band degeneracy points near the Fermi level, which display local linear dispersion in all directions \cite{Burkov2011, Wan2011}. Layered transition metal dichalcogenides (TMDCs) compounds $MX_{2}$ ($M$ = Mo, W; $X$ = S, Se, Te) with low symmetry crystal such as $1T'$ ($P_{21}/m$) and $T_d$ ($Pmn_{21}$) phases represent a fascinating class of WSM candidates \cite{Chiu2020, Fatemi2018, Singh2020, Kang2019, Zhou2019, Li2018}. These materials have shown promise for enabling exotic phenomena such as topological superconductivity \cite{Chiu2020, Fatemi2018}, the nonlinear Hall effect \cite{Singh2020, Kang2019}, anisotropic spin Hall transport \cite{Zhou2019}, and out-of-plane spin-orbit torque \cite{Li2018}. When scaled down to the two-dimensional (2D) monolayer (ML) limit, they transition from the bulk type-II WSM phase to the quantum spin Hall (QSH) state \cite{Tang2017, Chen2018, Fei2017}, where the topological gap can be easily tuned, for instance, through strain effect \cite{Zhao2020}, defect \cite{Muechler2020}, and application of an external electric fields \cite{Maximenko2022}.

The discovery of the QSH state in 2D TMDCs with low symmetry crystals ($1T'$ and $T_d$) has opened new avenues for spintronics and quantum metrology due to the ability of topologically protected states to transport spin information over long distances with minimal energy loss \cite{Tang2017, Chen2018, Fei2017}. The QSH state arises from strong SOC, which is fundamentally linked to the system's symmetries \cite{Kane2005}. However, even in systems with time-reversal symmetry, the lack of a well-defined spin conservation axis in the QSH state results in backscattering in edge states, undermining the ideal ballistic transport \cite{Sheng2006, Anders2010, Schmidt2012}. One way to overcome this obstacle is by promoting spin conservation through the design of systems that support persistent spin texture (PST) states—a formation of unidirectional spin configuration in momentum ($k$)-space, which extends spin lifetimes even in the presence of chemical and structural imperfections \cite{Bernevig2006, Schliemann2017, Kammermeier2020}. The PST state has previously been demonstrated in semiconductor quantum wells (QWs) where the Dresselhaus and Rashba SOC are of equal strength, or in [110]-oriented semiconductor QWs described by the [110] Dresselhaus model \cite{Bernevig2006}. Recently, a more robust PST state has been proposed by leveraging crystal symmetry, as demonstrated in several bulk ferroelectric \cite{Tao2018} and 2D systems \cite{Ji2022, Absor2019, Absor2022, Sasmito2021, Guo2023}. This PST effect presents exciting potential for spintronics, especially when combined with dissipationless chiral edge states.

Recently, both theoretical \cite{Garcia2020, Vila2021, Zhao2023} and experimental \cite{WenjinZ2021, Tan2021} studies have reported the observation of the PST-driven canted quantum spin Hall (QSH) effect in the $T_d$ phases of W(Mo)Te$_2$ monolayers (ML), highlighting a new potential use of topological 2D TMDCs materials in spintronics. In the $T_d$ W(Mo)Te$_2$ ML, the lack of inversion symmetry and the presence of multiple vertical mirror planes result in a unidirectional canted spin texture near the Fermi level \cite{Garcia2020, Vila2021, AbsorA2022}. This allows the topologically protected edge states to generate a quantized spin Hall conductivity plateau of $2e^2/h$ along the canting axis \cite{Garcia2020, Vila2021}. However, the $T_d$ phase is not the only structural form; WTe$_2$ ML also tends to favor the $1T'$ phase \cite{You2018, Zhou2016, Zhao2020}, which is energetically more stable. In the centrosymmetric $1T'$ phase, spin degeneracy persists throughout the electronic band structure, limiting its usefulness for spintronic applications. Additionally, the $1T'$ phase of the $MX_2$ ML has demonstrated a range of tunable properties in response to surface imperfections, such as defects \cite{Muechler2020, Absor2024} and Janus $MXX'$ ($M$ = Mo, W; $X\neq X'$= S, Se, Te) structures with surface alloying effect \cite{Joseph_2021, Tang2018, Varjovi2021, Li2020}, which can lead to phenomena like topological switching \cite{Muechler2020, Joseph_2021} and modulation of spin splitting and SOC parameters \cite{Absor2024}. Despite these advancements, the possibility of controlling the canted PST in the TMDCs $MX_{2}$ MLs by the Janus structures created by the surface alloying effect still remains unexplored. This could pave the way for designing controllable spin devices, enhancing the potential of dissipationless spintronics and quantum metrology.

In this paper, we employed first-principles density-functional theory (DFT) calculations to show the tunable spin splitting and canted PST in the 2D $1T'$ $MX_{2}$ MLs by utilizing chalcogen ($X$) elements of the ML surface to form highly stable Janus $1T'$ $MXX'$ ($M$ = Mo, W; $X\neq X'$= S, Se, Te) structure. Previously, stability of the Janus $1T'$ $MXX'$ MLs has been theoretically predicted \cite{Joseph_2021, Tang2018, Varjovi2021, Li2020}, and their experimental synthesis has also been reported \cite{Liu2024, Shahmanesh2021}. Consistent with previous report \cite{Joseph_2021, Tang2018, Varjovi2021, Li2020}, we have also confirmed the stability of the Janus $1T'$ $MXX'$ MLs by performing phonon dispersion analysis, ab initio molecular dynamics simulations, and formation energy calculations. By using the Janus $1T'$ WSTe ML as a representative example, we show that a strongly anisotropic spin-split bands displaying canted PST is observed in the electronic states near the Fermi level, which contrast sharply with those typically found in commonly studied PST materials \cite{Bernevig2006, Schliemann2017, Kammermeier2020, Tao2018, Ji2022, Absor2019, Absor2022, Sasmito2021, Guo2023}. We elucidated that these intricate spin splitting and spin textures are originated from the strong in-plane $p-d$ orbital coupling between the chalcogen atoms (Te and Se) and tungsten (W) atoms, induced by the lowering of the crystal's point group symmetry. The observed anisotropic spin splitting together with canted PST are also clarified  by a $\vec{k}\cdot\vec{p}$ model derived from symmetry analysis. More importantly, the spin-split states are highly responsive to surface imperfections caused by surface alloying effects, such as changes in the concentration of different chalcogen atoms on the ML surface. This suggests that the Janus $1T'$ $MXX'$ MLs offer a promising platform for future spintronic devices.
 
\section{Model and Computational Details}

We conducted DFT calculations using norm-conserving pseudopotentials and optimized pseudo-atomic localized basis functions \cite{Troullier}, as implemented in the OPENMX code \cite{Ozaki2003, Ozaki2005, Ozaki2004}. The generalized gradient approximation (GGA-PBE) by Perdew, Burke, and Ernzerhof was chosen for the exchange-correlation functional \cite{Perdew1996, Kohn1965}. The basis functions were expressed as a linear combination of multiple pseudo-atomic orbitals (PAOs) generated through a confinement scheme \cite{Ozaki2003, Ozaki2005, Ozaki2004}. Specifically, two $s$-, two $p$-, and two $d$-type numerical PAOs were employed. The FBZ integration was performed with a $12\times8\times1$ $k$-point mesh. To avoid artificial interactions between periodic images caused by boundary conditions, we applied a periodic slab model to the 2D ML systems, incorporating a vacuum layer of 25 \AA\ in the non-periodic direction. We optimized both the lattice and atomic positions until the Hellmann-Feynman forces on each atom were below $10^{-3}$ eV\AA, with an energy convergence threshold of $10^{-9}$ eV. 

To confirm the stability of the Janus $1T'$ $MXX'$ MLs, we calculate the formation energy, $E_{f}$, by using the following relation \cite{Joseph_2021, Tang2018},
\begin{equation}
\label{1}
E_{f}=\left(\frac{1}{n_{X}+n_{X'}+n_{M}}\right)\left[E_{MXX'}-\left(n_{X}E_{X}+n_{X'}E_{X'}+n_{M}E_{M}\right)\right],  
\end{equation}
where $E_{MXX'}$ is the total energy of Janus $1T'$ $MXX'$ MLs. $E_{X}$, $E_{X'}$, and $E_{M}$ are the chemical potential of
isolated $X$, $X'$, and $M$ atoms, respectively. $n_{X}$, $n_{X'}$, and $n_{M}$ are the number of $X$, $X'$, and $M$ atoms in the super cell or unit cell, respectively. Phonon dispersion band was used to evaluate the dynamical stability of the Janus $1T'$ $MXX'$ MLs calculated by using ALAMODE code \cite{Tadano2015} based on the force constants obtained from the OpenMX code calculations. The thermodynamic stability is evaluated by performing ab initio molecular dynamics (AIMD) simulations for 5 ps at a temperature of 600 K, controlled by a Nose-Hoover thermostat implemented in the openMX code.

The spin vector components ($S_{x}$, $S_{y}$, $S_{z}$) of the spin polarization along the reciprocal lattice vector $\vec{k}$ were determined by analyzing the spin density matrix. The spin density matrix, represented as $P_{\sigma \sigma^{'}}(\vec{k},\mu)$, is calculated from the spinor Bloch wave function, $\Psi^{\sigma}_{\mu}(\vec{r},\vec{k})$, using the following equation \cite{Kotaka_2013}, 
\begin{equation}
\begin{aligned}
\label{2}
P_{\sigma \sigma^{'}}(\vec{k},\mu)=\int \Psi^{\sigma}_{\mu}(\vec{r},\vec{k})\Psi^{\sigma^{'}}_{\mu}(\vec{r},\vec{k}) d\vec{r}\\
                                  = \sum_{n}\sum_{i,j}[c^{*}_{\sigma\mu i}c_{\sigma^{'}\mu j}S_{i,j}]e^{\vec{R}_{n}\cdot\vec{k}},\\
\end{aligned}
\end{equation}
where $\Psi^{\sigma}_{\mu}(\vec{r},\vec{k})$ is obtained after self-consistent is achieved in the DFT calculation. In Eq. (\ref{1}), $S_{ij}$ is the overlap integral of the $i$-th and $j$-th localized orbitals, $c_{\sigma\mu i(j)}$ is expansion coefficient, $\sigma$ ($\sigma^{'}$) is the spin index ($\uparrow$ or $\downarrow$), $\mu$ is the band index, and $\vec{R}_{n}$ is the $n$-th lattice vector.

\begin{figure*}
	\centering		
	\includegraphics[width=0.85\textwidth]{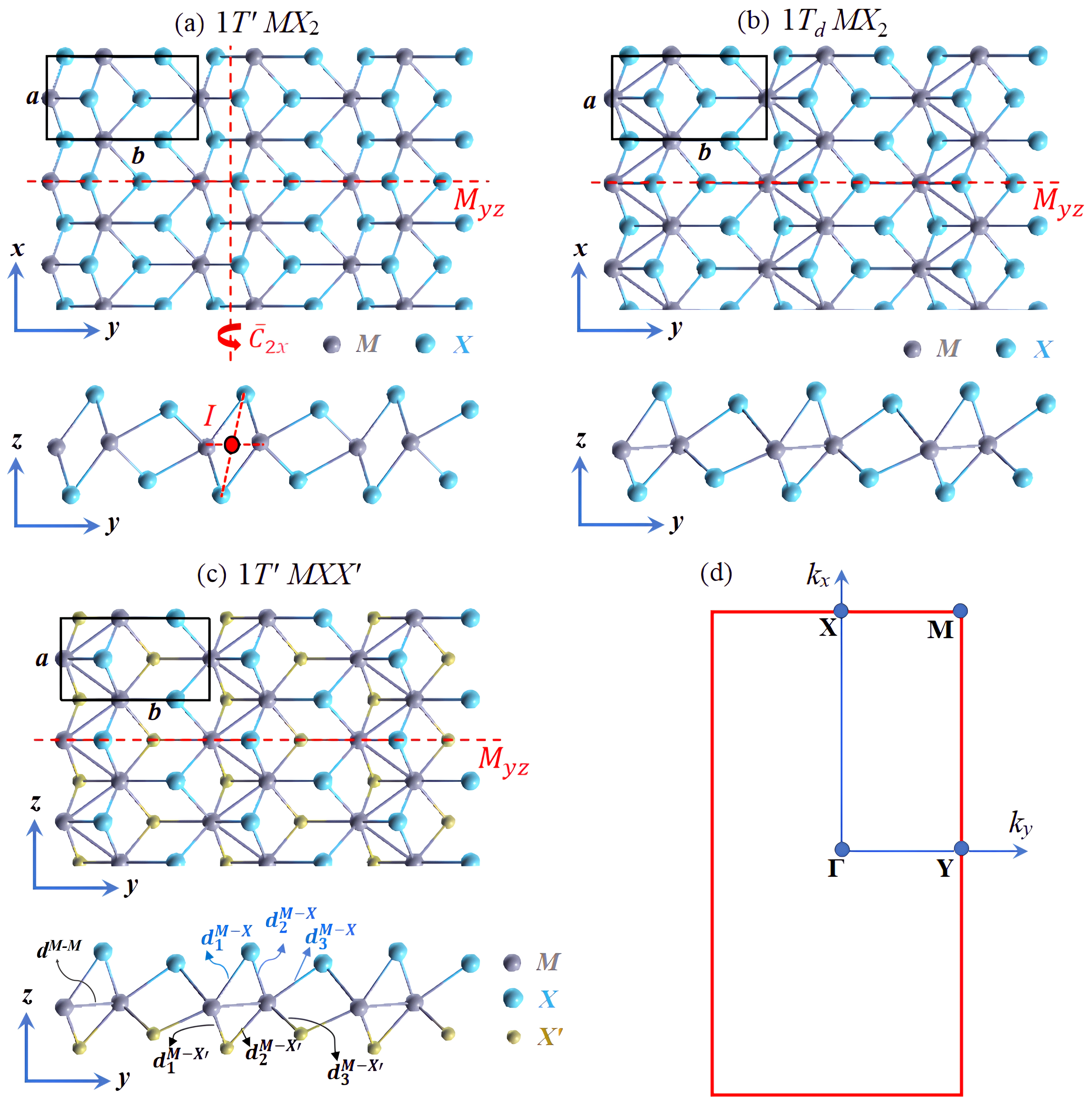}
	\caption{Atomic structures of (a) pristine $1T'$ $MX_{2}$ MLs, (b) pristine $1T_{d}$ $MX_{2}$ MLs, and (c) Janus $1T'$ $MXX'$ MLs ($M$ = Mo, W; $X \neq X'$ = S, Se, Te) are shown. The unit cell of each crystal is outlined by black lines and is characterized by the lattice parameters $a$ and $b$ in the $x$ and $y$ directions, respectively. The symmetry operations of the crystals, including inversion symmetry ($I$), mirror symmetry ($M_{yz}$) perpendicular to the $x$ axis, and two-fold screw rotation symmetry ($\bar{C}_{2x}$) around the $x$ axis, are indicated. The bond lengths between the $M$ and $X$ ($X'$) atoms—$d^{M-M}$, $d^{M-X}_{1}$, $d^{M-X}_{2}$, $d^{M-X}_{3}$, $d^{M-X'}_{1}$, $d^{M-X'}_{2}$, and $d^{M-X'}_{3}$—are schematically illustrated. The two-dimensional first Brillouin zone for both pristine and Janus MLs is shown, with time-reversal symmetry points such as $\Gamma$, as well as high-symmetry points $M$, $Y$, and $X$, highlighted.}
	\label{figure:Figure1}
\end{figure*}

\section{Results and Discussion}

The low-symmetry crystals of the 2D $MX_2$ MLs can be described by two possible structural phases, $1T'$ and $1T_{d}$, which arise due to differences in atomic arrangements and distortions [Fig. 1(a)-(b)]. The $1T'$ phase results from a distortion of the trigonal (octahedral) symmetry, leading to a monoclinic crystal system with a space group of $P2_{1}/m$ \cite{You2018, Zhou2016, Zhao2020, Absor2024}. This phase retains inversion symmetry due to the formation of zigzag chains of metal atoms ($M$) [Fig. 1(a)]. The $1T'$ $MX_2$ MLs belongs to the $C_{2h}$ point group, which is defined by the following symmetry operations: (i) identity operation $E$, (ii) inversion symmetry $I$, (iii) mirror symmetry $M_{yz}$ perpendicular to the $x$ axis, and (iv) a two-fold screw rotation symmetry $\bar{C}_{2x}$ along the $x$ axis, as shown in Fig. 1(a). The $1T_{d}$ phase of the $MX_2$ MLs represents a further distortion of the $1T'$ form \cite{Garcia2020, Vila2021, Zhao2023}, resulting in an orthorhombic crystal system characterized by a wave-like or dimerized arrangement of metal atoms ($M$) [Fig. 1(b)]. This phase retains mirror symmetry $M_{yz}$ but lacks both inversion symmetry $I$ and screw rotation symmetry $\bar{C}_{2x}$ [Fig. 1(b)]. Consequently, the $1T_{d}$ phase is non-centrosymmetric, with a $Pmn2_{1}$ space group and a lower point group symmetry of $C_s$. Our optimized structure calculations show that the $1T'$ phase has lower energy than the $1T_{d}$ phase [see Table S1 in the Supplementary Materials \cite{Supporting}], indicating that the $1T'$ phase is more favorable experimentally. Indeed, the $1T'$ phase of the WTe$_2$ ML has been experimentally reported \cite{Tang2017, Chen2018, Fei2017, Maximenko2022}.

\begin{table}[h]
\caption{The optimized structural parameters of the Janus $1T'$ $MXX'$ MLs such as in-plane lattice constants ($a$, $b$), bond length between $M$ atoms ($d^{M-M}$), and bond length between $M$ and $X$ ($X'$) atoms including $d^{M-X}_{1}$, $d^{M-X}_{2}$, $d^{M-X}_{3}$, $d^{M-X'}_{1}$, $d^{M-X'}_{2}$, and $d^{M-X'}_{3}$ as schematically shown in Fig. 1(c). All the structural parameters are measured in \AA. The formation energy $E_{f}$ calculated by Eq. (\ref{2}) is shown. Here, $E_{f}$ is measured in eV.} 
\centering 
\begin{tabular}{cc cc cc cc cc cc cc cc cc cc cc cc} 
\hline\hline 
 2D Materials && $a$  && $b$  && $d^{M-M}$  && $d^{M-X}_{1}$ && $d^{M-X}_{2}$ && $d^{M-X}_{3}$ && $d^{M-X'}_{1}$ && $d^{M-X'}_{2}$ && $d^{M-X'}_{3}$ && $E_{for}$ \\ 
\hline 
WSeTe        && 3.37 && 6.08 && 2.81  && 2.78 && 2.76  && 2.80 && 2.55 && 2.59 && 2.64 && -0.47\\
WSTe         && 3.33 && 6.01 && 2.82  && 2.78 && 2.76  && 2.80 && 2.43 && 2.46 && 2.51 && -0.67\\
WSSe         && 3.16 && 5.72 && 2.77  && 2.55 && 2.57  && 2.61 && 2.47 && 2.42 && 2.53 && -1.04\\
MoSeTe       && 3.36 && 6.08 && 2.83  && 2.77 && 2.75  && 2.80 && 2.54 && 2.57 && 2.63 && -0.58\\
MoSTe        && 3.28 && 6.06 && 2.85  && 2.76 && 2.74  && 2.81 && 2.41 && 2.46 && 2.49 && -0.50\\
MoSSe        && 3.11 && 5.74 && 2.77  && 2.56 && 2.53  && 2.61 && 2.40 && 2.45 && 2.47 && -0.75\\
\hline\hline 
\end{tabular}
\label{table:Table 1} 
\end{table}

By leveraging the chalcogen $X$ atoms on one surface of the $1T'$ $MX_2$ MLs, Janus $1T'$ $MXX'$ MLs can be formed \cite{Joseph_2021, Tang2018}. In this structure, both inversion symmetry $I$ and screw rotation symmetry $\bar{C}_{2x}$ are broken, as shown in Fig. 1(c). The optimized structural parameters for the Janus $1T'$ $MXX'$ MLs are summarized in Table I. Notably, the lattice constants ($a$, $b$) for the Janus $1T'$ $MXX'$ MLs decrease as the atomic numbers of the $M$ and $X,X'$ elements decrease. For instance, the WSeTe ML shows the largest values, with $a = 3.36$ \AA\ and $b = 6.08$ \AA, whereas the MoSSe ML has the smallest values, with $a = 3.11$ \AA\ and $b = 5.74$ \AA, consistent with previous studies \cite{Joseph_2021, Tang2018}. Furthermore, the significant difference between the $a$ and $b$ parameters in the crystal structure of the Janus $1T'$ $MXX'$ MLs suggests a highly anisotropic mechanical response when subjected to uniaxial strain along the $x$- and $y$-axes \cite{Varjovi2021}. Compared to pristine $1T'$ $MX_{2}$ MLs [see Table S1 in the Supplementary Materials \cite{Supporting}], the lattice constants ($a$, $b$) of the Janus structures are approximately the average of those of $1T'$ $MX_{2}$ and $1T'$ $MX'_{2}$ MLs. Additionally, the strong covalent bonds between the $M$ atoms in the Janus $1T'$ $MXX'$ MLs result in the substantial $M$-$M$ bond length ($d^{M-M}$), which slightly deviates from that of the pristine $1T'$ $MX_{2}$ and $1T'$ $MX'_{2}$ MLs. However, as shown in Table I, the bond lengths between the $M$ and $X$ ($X'$) atoms, $d^{M-X}$ ($d^{M-X'}$), decrease in Janus structures as the atomic number of the $M$ and $X$ ($X'$) elements decrease, reflecting the same trend observed in the lattice constants.

\begin{figure*}
	\centering		
	\includegraphics[width=1.0\textwidth]{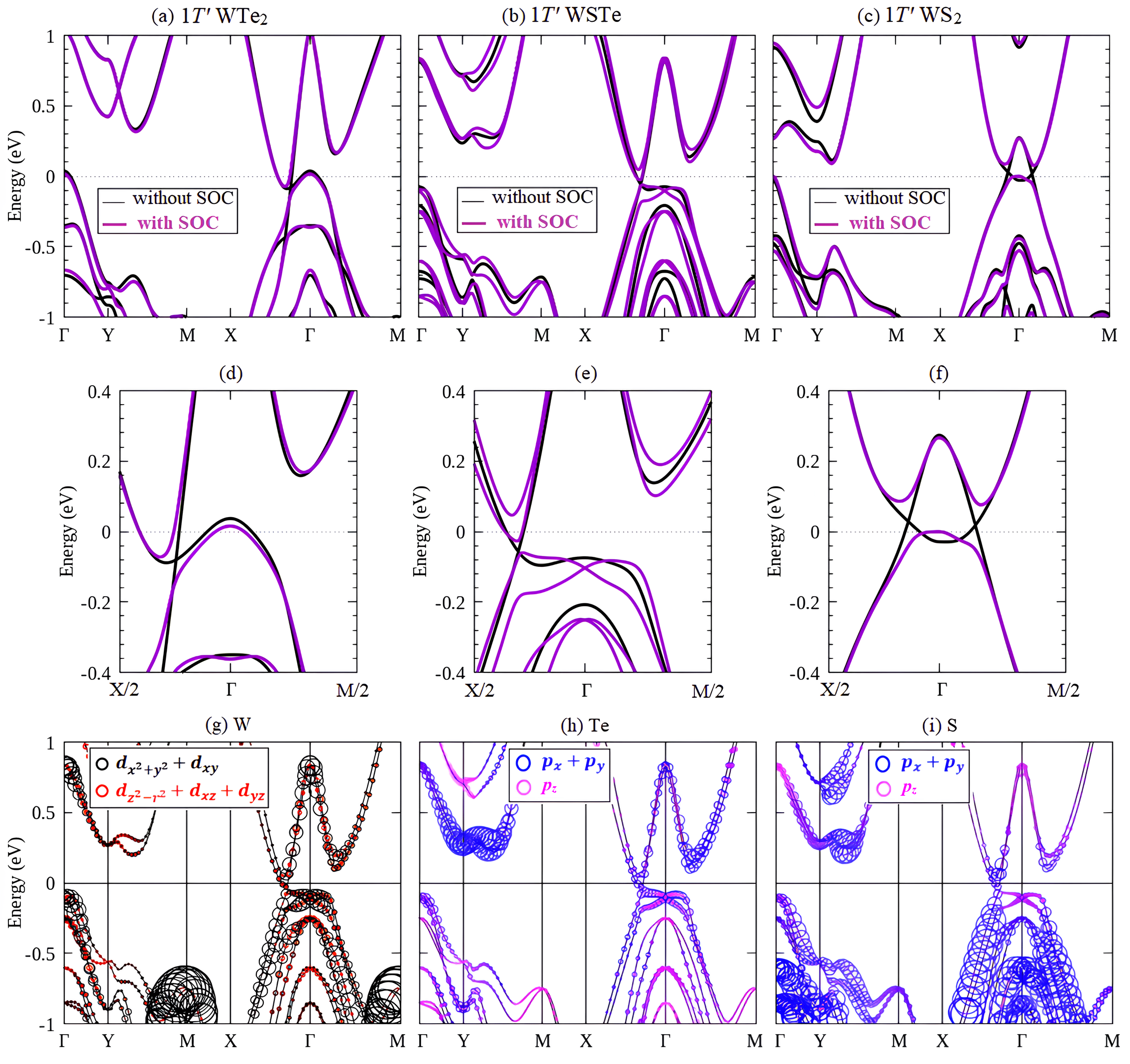}
	\caption{Electronic band structure along the $\Gamma-Y-M-X-\Gamma-M$ path of the FBZ is shown for (a) the pristine $1T'$ WTe$_{2}$ ML, (b) the Janus $1T'$ WSTe ML, and (c) the pristine $1T'$ WS$_{2}$ ML calculated without (black lines) and with (red lines) SOC. The band dispersion, emphasizing the spin-split bands along the $X/2-\Gamma-M/2$ direction at near the Fermi level, is presented for (d) the pristine $1T'$ WTe$_{2}$ ML, (e) the Janus $1T'$ WSTe ML, and (f) the pristine $1T'$ WS$_{2}$ ML. Orbital-resolved projected bands for the $1T'$ WSTe ML are displayed for (g) W, (h) Te, and (i) S atoms. The radii of the circles represent the magnitudes of the spectral weight of the specific orbitals contributing to the bands.}
	\label{figure:Figure2}
\end{figure*}

The stability of the Janus $1T'$ $MXX'$ MLs is evaluated by calculating the formation energy $E_f$. All the examined Janus $1T'$ $MXX'$ MLs show negative formation energies ($E_{f}<0$) [see Table I], suggesting that these structures are energetically favorable and could be experimentally realized. In fact, several Janus $1T'$ $MXY$ MLs such as $1T'$ WSSe ML \cite{Shahmanesh2021} and $1T'$ MoSSe ML \cite{Liu2024} have been experimentally reported. Additionally, the phonon dispersion bands, provided in Fig. S1 of the supplementary materials, further verify the stability of the Janus $1T'$ $MXX'$ MLs, which is consistent with previous reports \cite{Joseph_2021, Tang2018, Varjovi2021, Li2020}. The absence of imaginary frequencies in the phonon dispersion confirms the dynamic stability of the optimized Janus $1T'$ $MXX'$ structures. The thermodynamic stability of the Janus $1T'$ $MXX'$ MLs was further assessed using AIMD simulations at 600 K with a $3\times3\times1$ supercell. The results indicate that the different total energies of the Janus structures fluctuated smoothly with minimal variations. Furthermore, no significant structural breakdowns were observed after 1.5 ps, as shown in the snapshot of the $1T'$ $MXX'$ MLs in Fig. S2 of the supplementary materials \cite{Supporting}. These findings suggest that the Janus structures possess high thermal stability and are likely to be experimentally observable at temperatures beyond room temperature. Given that all the Janus systems exhibit similar structural symmetry, as well as comparable energetic, dynamic, and thermal stability, the subsequent discussion will focus on the $1T'$ WSTe ML as a representative example of Janus $1T'$ $MXX'$ MLs. 

Figs. 2(a)–(c) present the electronic band structure of the Janus $1T'$ WSTe ML in comparison to the pristine $1T'$ W$X_{2}$ ($X$ = Se, Te) MLs, calculated along a specific $\vec{k}$ path ($\Gamma-Y-M-X-\Gamma-M$) in the FBZ [Fig. 1(d)]. The electronic band structures for other Janus $1T'$ $MXX'$ MLs are shown in Fig. S3 of the Supplementary Materials \cite{Supporting}. In the absence of SOC, both the pristine $1T'$ W$X_{2}$ MLs and the Janus $1T'$ WSTe ML exhibit band crossings between the highest occupied state (HOS) and the lowest unoccupied state (LUS) at a specific $k$-point near the Fermi level along the $X-\Gamma$ line, forming a degenerate Dirac nodal point. This degenerate point occurs at -0.05 eV for the Janus $1T'$ WSTe ML [Fig. 2(e)], which is slightly higher in energy than the $1T'$ WTe${2}$ ML (-0.08 eV) but lower than the $1T'$ WS${2}$ ML (0.04 eV) [Fig. 2(f)]. When SOC is introduced, the degenerate Dirac point opens, rendering both the pristine $1T'$ W$X_{2}$ MLs and the Janus $1T'$ WSTe ML as 2D topological insulators \cite{Zhao2020, Muechler2020, Vila2021, Joseph_2021, Li2020}. However, due to the presence of both time-reversal symmetry and inversion symmetry in the pristine $1T'$ W$X_{2}$ MLs, all bands remain degenerate across the entire FBZ [Figs. 2(a) and 2(c); Figs. 2(d) and 2(f)]. In contrast, the absence of inversion symmetry in the Janus $1T'$ WSTe ML results in significant spin splitting of the bands due to SOC, particularly evident at the HOS and LUS along the $X-\Gamma-M$ line [Figs. 2(b) and 2(e)]. Orbital-resolved calculations confirm that the states near the Fermi level (HOS and LUS) in the Janus $1T'$ WSTe ML primarily arise from strong hybridization between the W-$d_{x^{2}+y^{2}}+d_{xy}$, Te-$p_{x}+p_{y}$, and S-$p_{x}+p_{y}$ orbitals [Figs. 2(g)–(i)].

\begin{figure*}
	\centering		
	\includegraphics[width=1.0\textwidth]{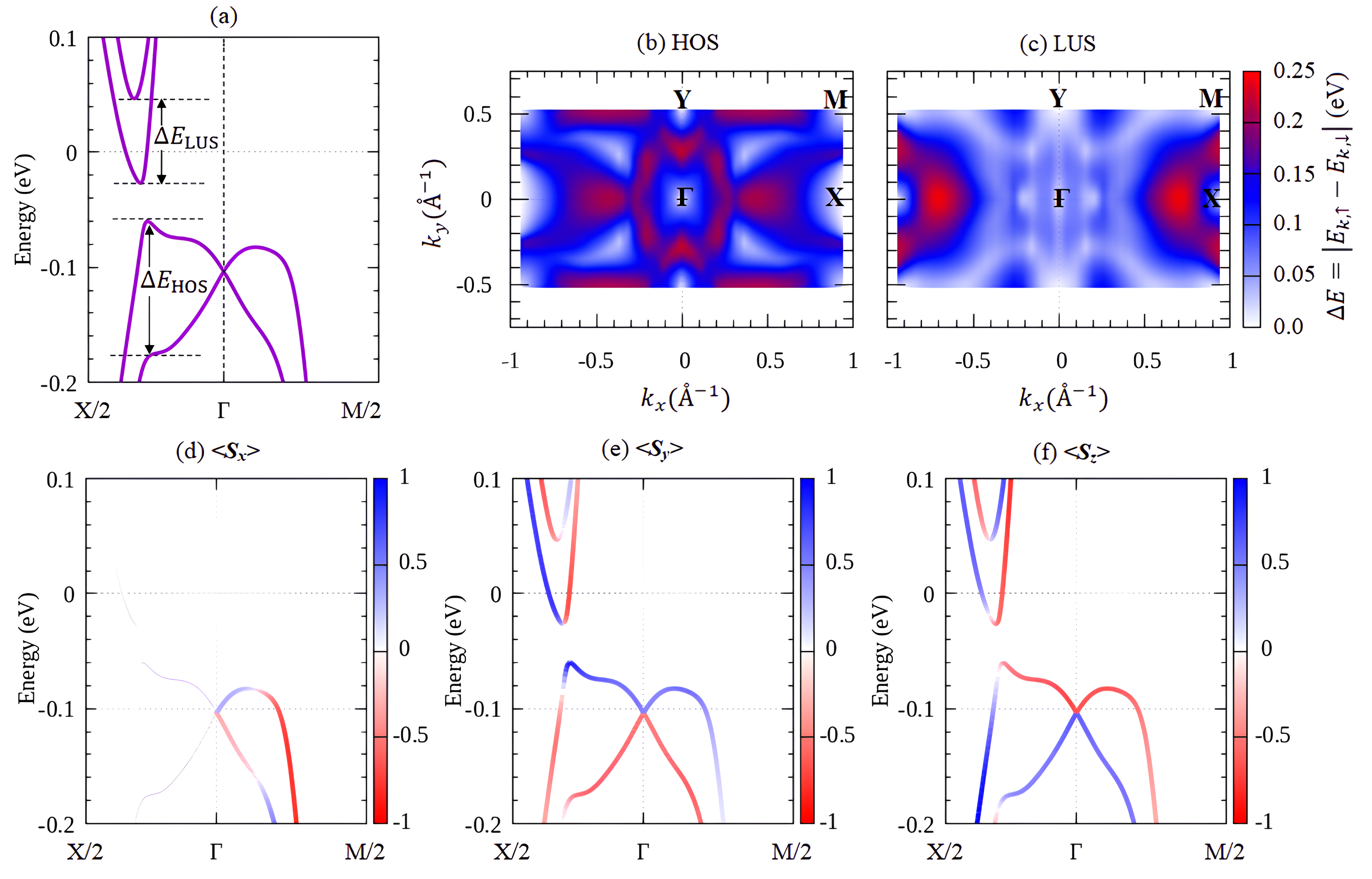}
	\caption{(a) The electronic band structure of the $1T'$ WSTe ML, calculated along the $X/2-\Gamma-M/2$ path near the Fermi level, highlighting the spin-splitting energies at the HOS ($\Delta E_{\texttt{HOS}}$) and LUS ($\Delta E_{\texttt{LUS}}$), is shown. (b)–(c) The spin-splitting energy mapped across the FBZ for the HOS and LUS bands, respectively, is displayed. The magnitude of the spin-splitting energy, $\Delta E$, is defined as $\Delta E = |E(k,\uparrow) - E(k,\downarrow)|$, where $E(k,\uparrow)$ and $E(k,\downarrow)$ are the energy bands of the HOS (or LUS) with up-spin and down-spin, respectively, and is represented by the color scale. (d)–(f) Spin-resolved bands near the Fermi level along the $X/2-\Gamma-M/2$ path, showing the expectation values of the spin components $\left\langle S_{x}\right\rangle$, $\left\langle S_{y}\right\rangle$, and $\left\langle S_{z}\right\rangle$, respectively, are also represented by the color scale.}
	\label{figure:Figure3}
\end{figure*}

\begin{table*}
\caption{The maximum spin-splitting energy for the HOS and LOS bands ($\Delta E_{\texttt{HOS}}$, $\Delta E_{\texttt{LUS}}$) measured in eV are shown. The calculated canting angles of the PST for the HOS and LOS bands along the $\Gamma-X$ direction ($\theta_{\texttt{HOS}}^{\Gamma-X}$,  $\theta_{\texttt{LUS}}^{\Gamma-X}$), determined through the relation $\theta_{\texttt{HOS},\texttt{LUS}}^{\Gamma-X}= \tan^{-1} \frac{\left\langle S_{z}\right\rangle_{k}}{\left\langle S_{y}\right\rangle_{k}}$, where the $\left\langle S_{y}\right\rangle$ and $\left\langle S_{z}\right\rangle$ are the expectation value of the $S_{y}$ and $S_{z}$ spin components in the spin-resolved bands at specific $\vec{k}$ points along the $\Gamma-X$ direction \cite{WenjinZ2021}, are given.} 
\centering 
\begin{tabular}{ccc  ccc  ccc   ccc   ccc} 
\hline\hline 
 2D Materials  &&&  $\Delta E_{\texttt{HOS}}$ (eV\AA)  &&& $\Delta E_{\texttt{LUS}}$ (eV\AA) &&&   $\theta_{\texttt{HOS}}^{\Gamma-X}$ &&& $\theta_{\texttt{LUS}}^{\Gamma-X}$ \\ 
\hline 
$1T'$ WSeTe ML             &&&   0.092    &&&   0.08    &&&  $32.29^{0}$   &&&  $17.74^{0}$   \\
 $1T'$ WSTe ML             &&&   0.14     &&&   0.10    &&&  $52.43^{0}$   &&&  $61.26^{0}$   \\
$1T'$ WSSe ML              &&&   0.04     &&&   0.01    &&&  $35.49^{0}$   &&&  $62.46^{0}$   \\
$1T'$ MoSeTe ML            &&&   0.03     &&&   0.04    &&&  $7.91^{0}$    &&&  $81.02^{0}$    \\
$1T'$ MoSTe ML             &&&   0.08     &&&   0.04    &&&  $12.29^{0}$   &&&  $70.09^{0}$     \\ 
$1T'$ MoSSe ML             &&&   0.02     &&&   0.03    &&&  $43.91^{0}$   &&&  $75.83^{0}$     \\
\hline\hline 
\end{tabular}
\label{table:Table 2} 
\end{table*}

Both the HOS and LUS bands are located near the Fermi level, indicating their significant impact on carrier transport properties. Examining the band dispersion of the Janus $1T'$ WSTe ML near the Fermi level, as shown in Fig. 3(a), the spin splitting displays an anisotropic behavior in both the HOS and LUS bands. This is evident in the calculated spin-splitting energy mapped across the entire FBZ region, as shown in Figs. 3(b)–3(c). The mapping reveals a highly anisotropic spin splitting, with notable differences in the spin-splitting energies between the $X-\Gamma$ and $\Gamma-M$ directions. The observed anisotropy in the HOS and LUS spin-splitting energies arises from the low crystal symmetry of the Janus $1T'$ WSTe ML, similar to the behavior previously reported in WTe$_{2}$ bilayers \cite{Absor2022}. Notably, a pronounced spin splitting is observed in both the HOS and LUS bands, with the computed maximum spin-splitting energies ($\Delta E_{\texttt{HOS}}$, $\Delta E_{\texttt{LUS}}$) for all Janus $1T'$ $MXX'$ MLs are listed in Table II. It is observed that the maximum spin-splitting energy is more significant in Janus structures with heavier elements and a greater degree of broken inversion symmetry. For the $1T'$ WSTe ML, we found that $\Delta E_{\texttt{HOS}} = 0.14$ eV and $\Delta E_{\texttt{LUS}} = 0.10$ eV, which are comparable to those reported for various 2D systems including TMDCs MLs with the $1H$ phase (0.03–0.50 eV) \cite{Zhu2011, Absor2016, Affandi2019}, their Janus counterparts (0.04–0.49 eV) \cite{Absor2017, Absor2018JJAP, Chakraborty2023}, and Janus group IV-V MLs (0.1-0.65 eV) \cite{ArifL_2023}. However, these values are significantly larger than those reported for Janus Bi$XY$ MLs (0-0.071 eV) \cite{Varjovi} and low-symmetry 2D $MX_{2}$ ML systems including $T_{d}$ WTe$_{2}$ and $T_{d}$ MoTe$_{2}$ MLs ($<1$ meV) \cite{Garcia2020, Zhao2023, Vila2021}. We emphasize here that the large spin-splitting energy observed in both the LUS and HOS bands of the present systems is originated from the strong in-plane $p-d$ coupling orbitals as confirmed by the calculated orbital-resolved bands shown in Figs. 2(g)–(i). These results align with the established understanding that strong in-plane $p-d$ orbital coupling plays a crucial role in inducing significant spin splitting, as previously reported in various 2D TMDC MLs \cite{Zhu2011, Absor2016, Affandi2019}. The substantial spin splitting observed in these systems highlights their potential for spintronic applications, even at room temperature \cite{Yaji2010}. 

To further explore the spin-splitting properties of the Janus $1T'$ WSTe ML in the HOS and LUS bands, we present the calculated spin-resolved bands along the $X-\Gamma-M$ directions, as shown in Figs. 3(d)-(f). The presence of the $M_{yz}$ mirror plane in the Janus $1T'$ WSTe ML leads to prominent $S_{y}$ and $S_{z}$ components of spin polarization in the spin-split bands along the $\Gamma-X$ direction, while the $S_{x}$ component remains nearly zero. However, the $S_{x}$ component becomes dominant in the spin-split bands along the $\Gamma-M$ direction. Since the $S_{y}$ and $S_{z}$ spin components are primarily responsible for the spin polarization in the HOS and LUS bands along the $\Gamma-X$ direction, a tilted unidirectional spin polarization is achieved within the $yz$ plane of the FBZ. This tilted spin polarization in momentum space results in a canted PST, which stands in contrast to the typical PST materials reported in prior studies \cite{Bernevig2006, Schliemann2017, Kammermeier2020, Tao2018, Ji2022, Absor2019, Absor2022, Sasmito2021, Guo2023}. Interestingly, a similar canted PST has been previously observed on the ZnO (10$\bar{1}$0) surface \cite{Absor_2015}, $1T'$ WTe$_{2}$ bilayer \cite{Absor2022}, and $T_{d}$ phase of TMDC $MX_{2}$ MLs such as $T_{d}$ WTe$_{2}$ ML \cite{Garcia2020, Zhao2023} and $T_{d}$ MoTe$_{2}$ ML \cite{Vila2021}. This canted PST is expected to induce a unidirectional canted spin-orbit field in momentum space, which helps protect spin coherence and leads to an exceptionally long spin lifetime \cite{Bernevig2006, Schliemann2017}. 

By evaluating the magnitudes of the $\left\langle S_{y}\right\rangle$ and $\left\langle S_{z}\right\rangle$ spin components in the spin-resolved bands at specific $\vec{k}$ points along the $\Gamma-X$ direction, the canting angle ($\theta_{\texttt{HOS},\texttt{LUS}}^{\vec{k}}$) of the PST along the $\Gamma-X$ direction can be estimated through the following relation, $\theta_{\texttt{HOS},\texttt{LUS}}^{\Gamma-X}= \tan^{-1} \frac{\left\langle S_{z}\right\rangle_{k}}{\left\langle S_{y}\right\rangle_{k}}$ \cite{WenjinZ2021}, and we find that the calculated $\theta_{\texttt{HOS},\texttt{LUS}}^{\Gamma-X}$ for the HOS and LUS bands are $52.43^{0}$ and $61.26^{0}$, respectively; see Table II. In particular, the calculated canting angle of PST for the LUS bands ($\theta_{\texttt{\texttt{LUS}}}^{\vec{k}}=61.26^{0}$) is larger than that reported for the $T_{d}$ phase of the WTe$_{2}$ ML ($\approx 56.0^{0}$) \cite{Garcia2020}. Notably, the observed canting angles of the PST in the $1T'$ $MXX'$ MLs could well-define a momentum-independent spin axis, which is crucial for determining the spin axis of the QSH state \cite{WenjinZ2021}. Therefore, the canted PST could link the spintronic and topological properties of Janus $1T'$ $MXX'$ MLs, offering potential for the development of highly efficient spintronic devices.

The observed spin splitting bands with canted PST in the Janus $1T'$ WSTe ML can be explained using an effective $\vec{k}\cdot\vec{p}$ Hamiltonian model, derived based on symmetry considerations. We apply the $\vec{k}\cdot\vec{p}$ model, previously used for generic $1T'$ and $T_{d}$ structures of $MX_{2}$ MLs, as it effectively captures key physical characteristics and provides an accurate description of their band structures in particularly for the bands at near the Fermi level \cite{Garcia2020, Vila2021, Zhao2023}. As illustrated in Figs. 2(a) and 2(c), the electronic states of pristine $1T'$ $MX_{2}$ MLs near the Fermi level are well-represented by two pairs of degenerate bands, inverted at the $\Gamma$ point. These band pairs correspond to the irreducible representations $\mathcal{A}_{g}$ ($d$ orbitals) and $\mathcal{B}_{u}$ ($p$ orbitals) of the $C_{2h}$ point group. When the Janus structures is introduced, both the inversion symmetry $I$ and the screw rotation $\bar{C}_{2x}$ are broken, resulting in two pairs of spin-split bands; see Fig. 2(b). By expanding the bands near the $\Gamma$ point, one can construct the following spin-full 4-bands $\vec{k}\cdot\vec{p}$ Hamiltonian \cite{Garcia2020, Vila2021, Zhao2023, Xie2020},
\begin{equation}
\label{3}
\mathcal{H}_{\Gamma}= \mathcal{H}_{0}+\mathcal{H}_{\texttt{SOC}} 
\end{equation} 
where the first term describes the Hamiltonian without the SOC: 
\begin{equation}
\label{4}
\mathcal{H}_{0}= \mathcal{A}\left(k_{x}^{2}+k_{y}^{2}\right)\sigma_{0}\tau_{0}+ \left[ \mathcal{B}\left(k_{x}^{2}+k_{y}^{2}\right)+\delta\right]\sigma_{0}\tau_{z}+ \beta k_{y}\sigma_{0}\tau_{y}.
\end{equation}
Here, $\tau_{i}$ ($\sigma_{i}$) with $i=x,y,z$ and $\tau_{0}$ ($\sigma_{0}$) are the Pauli matrices and identity matrix working in the orbital (spin) space, respectively. The parameters $\mathcal{A}$ and $\mathcal{B}$ are associated with the effective masses for both the HOS and LUS bands, while $\delta$ describes the degree of the band inversion at the center of the FBZ. The parameter $\beta$ quantifies the degree of crystalline anisotropy between $k_{x}$ and $k_{y}$ direction.  

The second term in Eq. (\ref{3}) represents the symmetry-allowed SOC Hamiltonian, $\mathcal{H}_{\texttt{SOC}}$. Given that our Janus $1T'$ $MXX'$ MLs belong to the $C_s$ point group, which is characterized by mirror symmetry $M_{yz}$ and time reversal symmetry $\mathcal{T}$ operations. Therefore, the symmetry-allowed $\mathcal{H}_{\texttt{SOC}}$ near the $\Gamma$ point can be derived by considering the mirror symmetry $M_{yz}=i\sigma_{x}\tau_{0}$ and the time reversal symmetry $\mathcal{T}=i\sigma_{y}\mathcal{K}$, where $\mathcal{T}^{2}=-1$ for the spinor and $\mathcal{K}$ represents complex conjugation. Taking the invariant term of of the $\mathcal{H}_{\texttt{SOC}}$ under the $M_{yz}$ and $\mathcal{T}$ operations; see Table III, we find that 
\begin{equation}
\label{5}
\mathcal{H}_{\texttt{SOC}}= \left(\alpha_{1} k_{x}\sigma_{y}+ \alpha_{2} k_{y}\sigma_{x}+ \alpha_{3} k_{x}\sigma_{z}\right) \tau_{x},
\end{equation}
where $\alpha_{1}$, $\alpha_{2}$, and $\alpha_{3}$ are the SOC parameters.

\begin{table*}
\caption{The transformations of ($\sigma_{x}$, $\sigma_{y}$, $\sigma_{z}$) and ($k_{x}$, $k_{y}$) with respect to the generators of the $C_{s}$ point group, including the mirror symmetry $M_{yz}=i\sigma_{x}\tau_{0}$ and the time reversal symmetry $\mathcal{T}=i\sigma_{y}\mathcal{K}$ operations, where $\mathcal{T}^{2}=-1$ for the spinor and $\mathcal{K}$ represents complex conjugation. The last column shows the terms which are invariant under point-group operation.} 
\begin{tabular}{cc cc cc cc} 
\hline\hline 
             \textbf{Operations}  && $(k_{x}, k_{y})$ && $(\sigma_{x}, \sigma_{y}, \sigma_{z})$ && \textbf{Invarian terms}  \\ 
\hline 
$M_{yz}=i\sigma_{x}\tau_{0}$    &&  $(-k_{x}, k_{y}, k_{z})$      &&   $(\sigma_{x}, -\sigma_{y}, -\sigma_{z})$       &&  $k_{n}^{m}k_{x}\sigma_{y}$, $k_{n}^{m}k_{x}\sigma_{z}$, $k_{n}^{m}k_{y}\sigma_{x}$; ($n=x,y$; $m=0,2$)\\			
$\mathcal{T}=i\sigma_{y}\mathcal{K}$   &&  $(-k_{x}, -k_{y}, -k_{z})$      &&  $(-\sigma_{x}, -\sigma_{y}, -\sigma_{z})$       &&  $k_{n}\sigma_{k}$; ($n=x,y$; $k=x,y,z$) \\
                  
\hline\hline 
\end{tabular}
\label{table:Table 3} 
\end{table*}

Solving the eigenvalue problem involving the Hamiltonian of Eqs. (\ref{3})-(\ref{5}), we obtain spin-dependent eigenvalues of the HOS and LUS bands [$E_{+}^{\texttt{HOS}}(k,\uparrow)$, $E_{-}^{\texttt{HOS}}(k,\downarrow)$, $E_{+}^{\texttt{LUS}}(k,\uparrow)$, $E_{-}^{\texttt{LUS}}(k,\downarrow)$] as follows,
\begin{equation}
\label{6}
E_{\pm}^{\texttt{HOS}}(k)= E_{0}^{\texttt{HOS}} \pm \sqrt{\left(\alpha_{1,\texttt{HOS}}^{2}+\alpha_{3,\texttt{HOS}}^{2}\right)k_{x}^{2}+\alpha_{2,\texttt{HOS}}^{2}k_{y}^{2}}  
\end{equation}
\begin{equation}
\label{7}
E_{\pm}^{\texttt{LUS}}(k)= E_{0}^{\texttt{LUS}} \pm \sqrt{\left(\alpha_{1,\texttt{LUS}}^{2}+\alpha_{3,\texttt{LUS}}^{2}\right)k_{x}^{2}+\alpha_{2,\texttt{LUS}}^{2}k_{y}^{2}}
\end{equation}
where $E_{0}^{\texttt{HOS}}$ and $E_{0}^{\texttt{LUS}}$ are the nearly free-electron energy for the HOS and LUS bands, respectively, which are given by the following equations:
\begin{equation}
\label{8}
E_{0}^{\texttt{HOS}}=\mathcal{A}_{\texttt{HOS}} \left(k_{x}^{2}+k_{y}^{2}\right) - \sqrt{\left[\mathcal{B}_{\texttt{HOS}}\left(k_{x}^{2}+k_{y}^{2}\right)+\delta_{\texttt{HOS}} \right]^{2} + \beta_{\texttt{HOS}}^{2}k_{y}^{2}}
\end{equation}
\begin{equation}
\label{9}
E_{0}^{\texttt{LUS}}=\mathcal{A}_{\texttt{LUS}} \left(k_{x}^{2}+k_{y}^{2}\right) + \sqrt{\left[\mathcal{B}_{\texttt{LUS}}\left(k_{x}^{2}+k_{y}^{2}\right)+\delta _{\texttt{LUS}}\right]^{2} + \beta_{\texttt{LUS}}^{2}k_{y}^{2}}.
\end{equation}
 
The normalized spinor wave functions for the spin-split pair bands, $\psi_{j}^{\pm}(k)$, where $j=(\texttt{HOS}, \texttt{LUS})$ is the index representing the HOS and LUS bands, are given by
\begin{equation}
\label{10}
\psi_{j}^{\pm}(k)=\frac{\psi_{0,j}^{\pm}(k)}{\sqrt{2\pi\left(\lambda_{\pm} +1\right)}}
\begin{pmatrix}
 \frac{\alpha_{1,j}k_{x}-i\alpha_{2,j}k_{y}}{\alpha_{3,j}k_{x} \mp E_{\texttt{SO}}^{j}} \\
		1
\end{pmatrix}
\end{equation}
where $\psi_{0,j}^{\pm}(k)$ is the pair wave functions of the free-electron,  $\lambda_{\pm}=\frac{\alpha_{1,j}^{2}k_{x}^{2}+\alpha_{2,j}^{2}k_{y}^{2}}{\left(\alpha_{3,j}k_{x} \mp E_{\texttt{SO}}^{j} \right)^{2}}$ and $E_{\texttt{SO}}^{j}=\sqrt{\left( \alpha_{1,j}^{2}+ \alpha_{3,j}^{2} \right)k_{x}^{2}+ \alpha_{2,j}^{2} k_{y}^{2} }$. The expectation value of the spin operator is obtained from $\left\langle \vec{S}\right\rangle_{j}^{\pm}=\frac{1}{2}\left\langle \psi_{j}^{\pm}|\vec{\sigma}|\psi_{j}^{\pm} \right\rangle$ resulting in
\begin{equation}
\label{11}
\left( \left\langle S_{x}\right\rangle, \left\langle S_{y}\right\rangle, \left\langle S_{z}\right\rangle  \right)_{j}^{\pm}=\pm\frac{1}{2E_{\texttt{SO}}^{j}} \left( \alpha_{2,j}k_{y}, \alpha_{1,j}k_{x}, \alpha_{3,j} k_{x} \right).
\end{equation}

From the spin-dependent energy dispersion of Eqs. (\ref{6})-(\ref{7}), the splitting energy of the spin-split bands, $\Delta E_{j} = |E_{j}(k,\uparrow) - E_{j}(k,\downarrow)|$, can be written as
\begin{equation}
\label{12}
\Delta E_{j} (k)=2 \sqrt{ \alpha_{(1,3),j}^{2} k_{x}^{2} + \alpha_{2,j}^{2} k_{y}^{2}},
\end{equation}
 where $\alpha_{(1,3),j}=\sqrt{\alpha_{1,j}^{2}+\alpha_{3,j}^{2}}$.

\begin{table*}
\caption{The calculated SOC parameters ($\alpha_{1,j}$, $\alpha_{2,j}$, $\alpha_{3,j}$) for the HOS and LUS bands in the Janus $1T'$ $MXX'$ MLs, measured in eV\AA, are shown. For comparison, the calculated SOC parameters for several selected low-symmetry ($T_d$ and $1T'$) 2D TMDC MLs and 2D Janus systems from previously reported calculations are also presented.} 
\centering 
\begin{tabular}{ccc  ccc  ccc  ccc  ccc  ccc   ccc  ccc } 
\hline\hline 
 2D Materials  &&&  $\alpha_{1,\texttt{HOS}}$ &&& $\alpha_{2,\texttt{HOS}}$ &&&   $\alpha_{3,\texttt{HOS}}$ &&&  $\alpha_{1,\texttt{LUS}}$ &&& $\alpha_{2,\texttt{LUS}}$ &&& $\alpha_{3,\texttt{LUS}}$ &&& Ref. \\ 
\hline 
Janus $1T'$ WSeTe ML             &&&   1.37     &&&   0.29   &&&  0.87   &&&  2.02  &&&  0.58    &&&   0.65   &&&  This work \\
Janus $1T'$ WSTe ML             &&&   0.78     &&&   0.15   &&&  1.02   &&&  0.15  &&&  0.07    &&&   0.28   &&&  This work \\
Janus $1T'$ WSSe ML              &&&   0.84     &&&   0.08   &&&  0.60   &&&  0.14  &&&  0.04    &&&   0.27   &&&  This work \\
Janus $1T'$ MoSeTe ML            &&&   0.24     &&&   0.05   &&&  0.03   &&&  0.08  &&&  0.03    &&&   0.48   &&&  This work \\
Janus $1T'$ MoSTe ML             &&&   1.54     &&&   0.15   &&&  0.34   &&&  0.22  &&&  0.10    &&&   0.60   &&&  This work \\ 
Janus $1T'$ MoSSe ML             &&&   0.43     &&&   0.01   &&&  0.41   &&&  0.15  &&&  0.03    &&&   0.57   &&&  This work \\
$T_{d}$ WTe$_{2}$ ML       &&&            &&&          &&&         &&&  0.059 &&&  0.078   &&&  0.116   &&&  Ref. \cite{Garcia2020} \\
                           &&&            &&&          &&&         &&&  0.072 &&&  0.084   &&&  0.121   &&&  Ref. \cite{Zhao2023}\\
$T_{d}$ MoTe$_{2}$ ML      &&&            &&&          &&&         &&&  0.095 &&&  0.159   &&&  0.089   &&&  Ref. \cite{Vila2021} \\
Janus $1H$ WSSe ML         &&&            &&&          &&&         &&&  0.48  &&&          &&&          &&&  Ref.\cite{Absor2018JJAP} \\
Janus $M$SSe ($M$=W, Mo) MLs &&&          &&&          &&&        &&&  0.004-0.17 &&&     &&&         &&&  Ref.\cite{Chakraborty2023} \\
Janus group IV-V MLs       &&&            &&&          &&&         &&&  0.08-1.53 &&&      &&&          &&&  Ref.\cite{ArifL_2023} \\
Janus Bi$XY$ MLs           &&&            &&&          &&&         &&&  0-1.98&&&        &&&          &&&  Ref.\cite{Varjovi} \\
Janus MXenes W$_{2}$CO$X$  &&&            &&&          &&&         &&&  0.45-0.85&&&       &&&          &&&  Ref.\cite{Arjyama2024} \\
\hline\hline 
\end{tabular}
\label{table:Table 4} 
\end{table*}

As demonstrated by Eq. (\ref{12}), the anisotropic spin splitting in both the HOS and LUS bands is influenced by the strength of the SOC parameters ($\alpha_{1,j}$, $\alpha_{2,j}$, $\alpha_{3,j}$). These parameters can be determined by analyzing the spin-split bands along the $\Gamma-X$ and $\Gamma-M$ directions. In the case of band splitting along the $\Gamma-X$ ($k_{x}$) direction, Eq. (\ref{12}) reduces to $\Delta E_{j}^{\Gamma-X} = 2 \alpha_{(1,3),j} k_{x}$. Consequently, $\alpha_{(1,3),j}$ can be extracted by fitting $\Delta E_{j}^{\Gamma-X}$ to the spin-splitting energy obtained from the DFT band dispersions along the $\Gamma-X$ direction. Moreover, Eq. (\ref{11}) allows for the determination of the ratio $(\alpha_{1,j}/\alpha_{3,j})$ by evaluating the in-plane $\left\langle S_{y} \right\rangle$ and out-of-plane $\left\langle S_{z} \right\rangle$ spin components for the HOS and LUS bands along the $\Gamma-X$ direction. This ratio is then used to derive $\alpha_{1,j}$ and $\alpha_{3,j}$ from the relation $\alpha_{(1,3),j} = \sqrt{{\alpha_{1,j}}^{2} + {\alpha_{3,j}}^{2}}$. Finally, once $\alpha_{1,j}$ and $\alpha_{3,j}$ are determined, $\alpha_{2,j}$ can be obtained by fitting the spin-splitting energy of the DFT band along the $\Gamma-M$ direction to Eq. (\ref{12}). 

The calculated SOC parameters ($\alpha_{1,j}$, $\alpha_{2,j}$, $\alpha_{3,j}$) for all Janus $1T'$ $MXX'$ MLs are summarized in Table IV, where they are compared with previously reported values for selected low-symmetry 2D $T_d$ $MX_2$ MLs and various 2D Janus ML systems. In general, the SOC parameters for the HOS bands in Janus $1T'$ $MXX'$ MLs are larger than those of the LUS bands, with the exception of the Janus $1T'$ WSeTe ML. Specifically, for the LUS band splitting, the SOC parameters ($\alpha_{1,\texttt{LUS}}$, $\alpha_{2,\texttt{LUS}}$, $\alpha_{3,\texttt{LUS}}$) are comparable to those found in various 2D Janus systems, such as Janus $M$SSe MLs \cite{Absor2018JJAP, Chakraborty2023}, Janus group IV-V MLs \cite{ArifL_2023}, Janus Bi$XY$ MLs \cite{Varjovi}, and Janus MXenes W${2}$CO$X$ MLs\cite{Arjyama2024}. However, these values are significantly larger than those reported for the $T_d$ phase of $MX_2$ MLs, including $T_d$ WTe$_{2}$ ML \cite{Garcia2020, Zhao2023} and $T_d$ MoTe$_{2}$ ML \cite{Vila2021}. Additionally, a noticeable variation in SOC parameters is observed across all Janus $1T'$ $MXX'$ MLs, indicating a strong spin-splitting anisotropy, consistent with the spin-splitting energy trends shown in Figs. 3(b)–3(c). Furthermore, the distinct values of $\alpha_{1,j}$ and $\alpha_{3,j}$ are expected to influence the spin polarization components $\left\langle S_{y} \right\rangle$ and $\left\langle S_{z} \right\rangle$ for both HOS and LUS bands along the $\Gamma-X$ direction. For instance, in the case of the $1T'$ WSTe ML, $\alpha_{3,\texttt{HOS}}$ (1.02 eV\AA) [$\alpha_{3,\texttt{LUS}}$ (0.28 eV\AA)] is substantially larger than $\alpha_{1,\texttt{HOS}}$ (0.78 eV\AA) [$\alpha_{1,\texttt{LUS}}$ (0.15 eV\AA)], leading to a substantial difference in the magnitudes of $\left\langle S_{y} \right\rangle$ and $\left\langle S_{z} \right\rangle$ for the HOS [LUS] bands along the $\Gamma-X$ direction. This difference results in a canted PST, aligning with the spin-resolved band structures shown in Figs. 3(e)–3(f).

\begin{figure*}
	\centering		
	\includegraphics[width=1.0\textwidth]{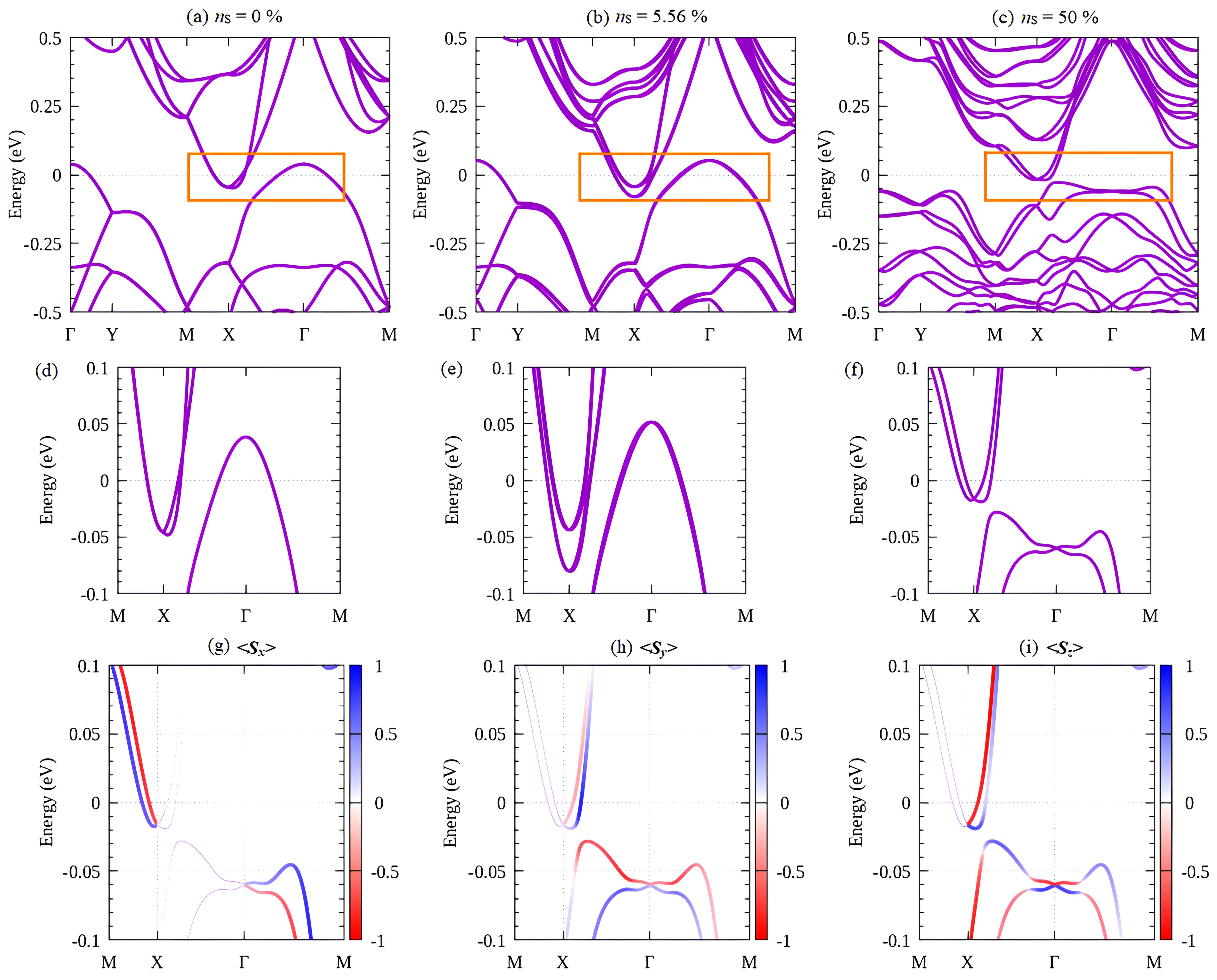}
	\caption{Concentration-dependent of the electronic band structure along the $\Gamma-Y-M-X-\Gamma-M$ path in the FBZ calculated with the SOC is shown for (a) $n_{\texttt{S}}=0$\%, (b) $n_{\texttt{S}}=5.56$\%, (c) $n_{\texttt{S}}=50$\%. The change of the spin-splitting bands at the the HUS and LUS close to the Fermi level are highlighted for the pristine ($n_{\texttt{S}}=0$\%) and alloy systems ($n_{\texttt{S}}=5.67$\%, $n_{\texttt{S}}=50$\%) in (d)-(f), respectively. (g)-(i) Spin-resolved bands near the Fermi level along the $M-X-\Gamma-M$ direction for the $1T'$ WSTe ML alloy having $n_{\texttt{S}}=50$ \% of the S atoms on the top surface ML, showing the expectation values of the spin components $\left\langle S_{x}\right\rangle$, $\left\langle S_{y}\right\rangle$, and $\left\langle S_{z}\right\rangle$, respectively, are also represented by the color scale. }
	\label{figure:Figure4}
\end{figure*}

We have discovered that Janus $1T'$ $MXX'$ MLs exhibit anisotropic spin splitting with canted PST, suggesting their potential for spintronic applications. Typically, these Janus $1T'$ $MXX'$ MLs can be engineered by substituting chalcogen atoms, which disrupts the symmetry of originally symmetric systems.  Previously, the Janus structure of MoSSe ML in the $1H$ phase has been successfully synthesized either via a two-step process involving H$_{2}$ plasma stripping and thermal selenization \cite{Lu2017}, which replaces the whole layer of S atoms on either side of MoS$_{2}$ ML with Se atoms, or by controlled sulfurization of MoSe$_{2}$ ML \cite{Zhang2017}. More recently, the successful synthesis of Janus structures in the form of well-dispersed $1T'$ WSe$_{2x}$S$_{2-2x}$ ML alloys has also been reported using a colloidal growth method \cite{Shahmanesh2021}. In this process, exchange of the chalcogen atoms (Se and S atoms) occurs in pure $1T'$ WSe$_{2}$ ML and can be controlled by the chalcogen precursors, enabling selective modification of some chalcogen atoms on one surface layer to create the Janus $1T'$ WSe$_{2x}$S$_{2-2x}$ ML alloys. This surface alloying effect can also be extendable to other Janus $1T'$ $MXX'$ MLs, where tuning the content of the chalcogen substitutions (S, Se, and Te) is expected being sensitively affects their electronic and spintronic properties.  

To ensure discussion aiming to explore the significant role of the surface alloying effect in the spin-splitting properties of the Janus $1T'$ $MXX'$ MLs, we show in Fig. 4 the evolution of the electronic and spin-resolved bands of the Janus $1T'$ WSTe ML alloys under varying concentrations of sulfur (S) substitutions on the upper ML surface. Referring to the total number of Te atoms in the ($3\times3\times1$) supercell structure of the pristine $1T'$ WTe$_{2}$ ML, the concentration of sulfur S atoms ($n_{\texttt{S}}$) is expressed as a percentage relative to the number of tellurium Te atoms on the top ML surface. When $n_{\texttt{S}}=0$ \%, this corresponds to a pure $1T'$ WTe$_{2}$ ML, whereas a Janus $1T'$ WSTe ML is achieved when all the Te atoms on the top surface are replaced by S atoms ($n_{\texttt{S}}=100$ \%). In its original form with $n_{\texttt{S}}=0$\%, the supercell structure of $1T'$ WTe$_{2}$ ML retains centrosymmetric nature of the $C_{2h}$ point group, resulting in degenerate bands across the Brillouin zone; see Figs. 4(a) and 4(d).  When sulfur (S) atoms are substituted on the ML surface, both inversion symmetry ($I$) and screw rotation symmetry ($\bar{C}_{2x}$) in the $1T'$ WTe$_{2}$ ML are lifted, leading to the anticipated emergence of the spin-splitting bands, in which the magnitude of the spin splitting varies based on the level of the S substitution. As illustrated in Figs. 4(b)-4(e) and 4(c)-4(f), the $1T'$ WSTe ML alloys demonstrate substantial spin-splitting, with its magnitude increasing as the S concentration $n_{\texttt{S}}$ rises. For instant, the strong enhancement of the spin splitting is visible on the HOS and LUS bands along the $M-X-\Gamma-M$ direction when $n_{\texttt{S}}$ is raised from 5.56\% to 50\%; see Figs. 4(e)-4(f). Additionally, our results show that this concentration-dependent spin splitting in the $1T'$ WSTe ML alloys also sustains a canted PST, as demonstrated by the calculated spin-resolved bands along the $\Gamma-X$ direction for $n_{\texttt{S}}=50$\% in Figs. 4(g)-4(i). This concentration-dependent spin splitting, while maintaining the canted PST in Janus $1T'$ $MXX'$ ML alloys, could open new avenues for tuning spintronic functionality via surface alloying effects.

\begin{figure*}
	\centering		
	\includegraphics[width=0.85\textwidth]{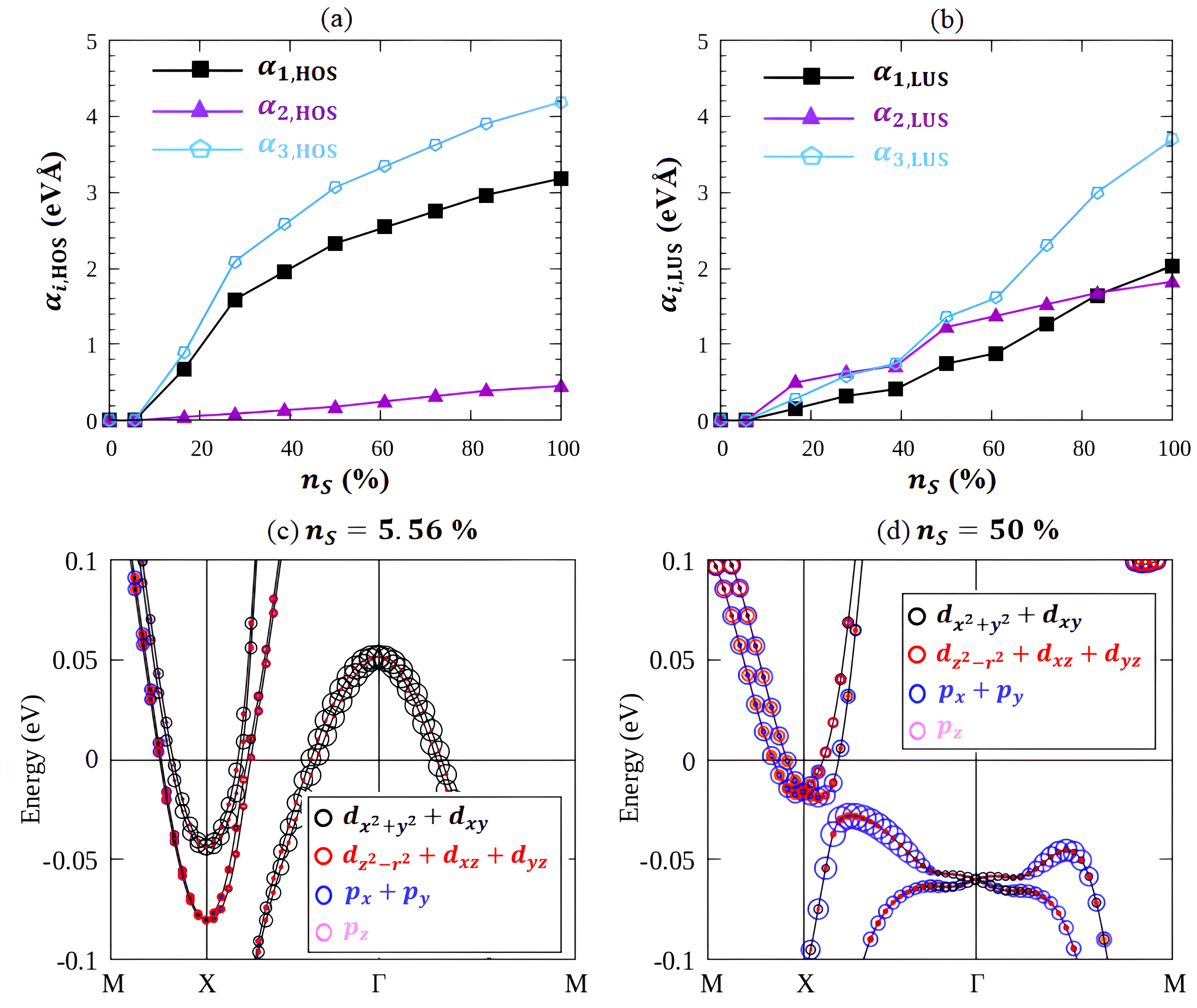}
	\caption{(a)-(b) Concentration-dependent of the S substitution in the Janus $1T'$ WSTe ML alloy on the SOC parameters for the HOS band around the $\Gamma$ point ($\alpha_{1,\text{HOS}}$, $\alpha_{2,\text{HOS}}$, $\alpha_{3,\text{HOS}}$) and LUS band around the $X$ point ($\alpha_{1,\text{LUS}}$, $\alpha_{2,\text{LUS}}$, $\alpha_{3,\text{LUS}}$), respectively, are shown. (c)-(d) Orbital-resolved projected bands near the Fermi level for the $1T'$ WSTe ML alloy with $n_{\texttt{S}}=5.56$\% and $n_{\texttt{S}}=50$\%, respectively. The radii of the circles represent the magnitudes of the spectral weight of the specific orbitals contributing to the bands.}
	\label{figure:Figure5}
\end{figure*}

The impact of chalcogen atom substitution during the surface alloying formation of the Janus $1T'$ $MXX'$ MLs on the spin-splitting properties is further quantified by evaluating the SOC parameters. In our Janus $1T'$ WSTe ML alloys, Figs. 4(e) and 4(f) show that the HOS bands are located close to the $\Gamma$ point, while the LUS bands are near the $X$ point. Given that both the $X$ and $\Gamma$ points in these alloys share the same little point group symmetry of the wave vector $\vec{k}$, maintaining only the mirror $M_{yz}$ plane, the symmetry-allowed SOC Hamiltonian from Eq. (\ref{5}) is applicable. The calculated SOC parameters for the HOS band around the $\Gamma$ point ($\alpha_{1,\text{HOS}}$, $\alpha_{2,\text{HOS}}$, $\alpha_{3,\text{HOS}}$) and the LUS band around the $X$ point ($\alpha_{1,\text{LUS}}$, $\alpha_{2,\text{LUS}}$, $\alpha_{3,\text{LUS}}$) are presented in Figs. 5(a) and 5(b), respectively, for various sulfur concentrations, $n_{\text{S}}$. This analysis reveals that all the SOC parameters exhibit a notable increasing trend with progressive S substitution. This trend can be understood by examining the hybridization of states through orbital-resolved calculations for the HOS and LUS bands of the Janus $1T'$ WSTe ML alloys at $n_{\text{S}}=5.56\%$ and $n_{\text{S}}=50\%$. As demonstrated in Figs. 5(c) and 5(d), an increase in in-plane $p$-$d$ orbital coupling is observed for the HOS and LUS bands along the $M$-$X$-$\Gamma$-$Y$ direction when the S concentration increases from $n_{\text{S}}=5.56\%$ to $n_{\text{S}}=50\%$, resulting in a significant enhancement of the spin splitting and SOC parameter values. The significant role of in-plane $p$-$d$ coupling orbitals in the increased spin splitting in the Janus $1T'$ WSTe ML alloys aligns with the widely studied spin-splitting phenomena in general 2D Janus TMDC systems with the $1H$ phase \cite{Absor2017, Absor2018JJAP}. Notably, the partial substitution of chalcogen atoms in $1T'$ $MX_{2}$ MLs through surface alloying can effectively amplify the spin-splitting properties of Janus $1T'$ $MXX'$ MLs, suggesting promising spintronic applications as experimental realization becomes feasible.

Before concluding, we would like to discuss the potential of the observed canted PST in the Janus $1T'$ $MXX'$ MLs as a platform for topological nanodevices aimed at spin transport applications. Notably, recent theoretical \cite{Garcia2020, Vila2021, Zhao2023} and experimental \cite{WenjinZ2021, Tan2021} evidences of a PST-driven canted QSH effect in low symmetric ($T_{d}$) W(Mo)Te$_{2}$ ML introduces a promising direction for topological materials in spintronics. Here, the topologically protected edge states inherit canted spin polarization from the 2D bulk bands, resulting in a quantized spin Hall conductivity plateau of $2e^{2}/h$ along the canting axis. The spin polarization generated by the canted QSH effect can exert an out-of-plane anti damping torque in magnets with perpendicular magnetic anisotropy \cite{Stiehl2019, MacNeill2017}, which are essential for next-generation, high-density spintronic applications. Given that the canted PST in the Janus $1T'$ $MXX'$ MLs shows larger spin-splitting energy and SOC parameters than $T_{d}$ W(Mo)Te$_{2}$ ML, it is expected that a more pronounced canted QSH effect should be experimentally observable. Additionally, a QSH effect with a canting angle modulated by the SOC parameter ratio could enable electrically tunable, dissipation-free spin currents with adjustable spin orientation, even without magnetic fields \cite{Garcia2022}. Thus, our results suggest a new type of gate-tunable spin-based device, topologically shielded against disorder and relevant for advancing topological spintronics.

\section{Conclussion}

We systematically conducted first-principles DFT calculations supplemented by $\vec{k}\cdot\vec{p}$ analysis to explore the spin-splitting characteristics of Janus $1T'$ $MXX'$ MLs (where $M$ = Mo, W, and $X \neq X'$ = S, Se, Te). Our findings reveal that the stable Janus $1T'$ $MXX'$ ML structure can be achieved by introducing different chalcogen ($X$) elements on the ML surface of $1T'$ $MX_{2}$ MLs. This stability is further confirmed through phonon band dispersion analysis, AIMD simulations, and formation energy calculations. By using Janus $1T'$ WSTe ML as a representative case, we have demonstrated the presence of strongly anisotropic spin-split bands, with maximum spin splittings of 0.14 eV and 0.10 eV occurring at the HOS and LUS bands, respectively. These significant band splittings give rise to the canted PST in the electronic states near the Fermi level, which exhibit distinct characteristics compared to conventional PST materials \cite{Bernevig2006, Schliemann2017, Kammermeier2020, Tao2018, Ji2022, Absor2019, Absor2022, Sasmito2021, Guo2023}. We attribute this complex spin splitting and spin texture to the strong in-plane $p-d$ orbital interaction between the chalcogen atoms $X$($X$ = Te and Se) and transition metal atoms $M$ ($M$ = W, Mo), driven by the lowered point group symmetry of the crystal. The observed anisotropic spin splitting and canted PST are also supported by a $\vec{k}\cdot\vec{p}$ model derived from symmetry analysis. Notably, these spin-split states show high sensitivity to surface imperfections, such as variations in chalcogen atom concentration due to surface alloying effects, indicating that Janus $1T'$ $MXX'$ MLs hold significant promise as platforms for next-generation spintronic devices.

Given the anisotropic spin splitting with canted PST observed in our study, it is reasonable to expect that this effect could also occur in other 2D Janus materials derived from pristine 2D structures with a $1T'$ phase. The asymmetry in the in-plane mirror operation $M_{yz}$, combined with the non-centrosymmetric nature of the crystals, serves as the primary driver of this phenomenon. Our symmetry analysis indicates that various 2D systems with a $1T'$ structure could support Janus configurations exhibiting spin-split bands with canted PST. These include 2D Janus structures based on $1T'$ $M$Si$_{2}Z_{4}$ ($M$ = Mo or W, $Z$ = P or As) monolayers (MLs) \cite{Islam2022} and $1T'$ $M$Si$_{2}$N$_{4}$ ($M$ = Ru or Os) MLs \cite{Varjovi2024}. As a result, our findings are expected to stimulate further theoretical and experimental research to identify new 2D materials exhibiting anisotropic spin splitting with canted PST, potentially advancing future spintronic applications.

\begin{acknowledgments}

This work was supported by the Academic of Excellence (AE) Program supported by Gadjah Mada University (No. 6526/UN1.P1/PT.01.03/2024). The computation in this research was performed using the computer facilities at Gadjah Mada University. 
\end{acknowledgments}

\bibliography{Reference1}


\end{document}